# Ultra-low Power Sensor Devices for Monitoring Physical Activity and Respiratory Frequency in Farmed Fish

Juan Antonio Martos-Sitcha[1,2], Javier Sosa[3], Dailos Ramos-Valido[3], Francisco Javier Bravo[4], Cristina Carmona-Duarte[5], Henrique Leonel Gomes[6], Josep Àlvar Calduch-Giner[1], Enric Cabruja[4], Aurelio Vega[3], Miguel Ángel Ferrer[5], Manuel Lozano[4], Juan Antonio Montiel-Nelson[3], Juan Manuel Afonso[7], Jaume Pérez-Sánchez[1,*]

[1]Nutrigenomics and Fish Growth Endocrinology Group, Institute of Aquaculture Torre de la Sal, Consejo Superior de Investigaciones Científicas (CSIC), 12595 Ribera de Cabanes, Castellón, Spain; [2]Department of Biology, Faculty of Marine and Environmental Sciences, Instituto Universitario de Investigación Marina (INMAR), Campus de Excelencia Internacional del Mar (CEI-MAR), University of Cádiz, 11519 Puerto Real, Cádiz, Spain. [3]Institute for Applied Microelectronics (IUMA), University of Las Palmas de Gran Canaria, Las Palmas, Spain. [4]Institute of Microelectronics of Barcelona (IMB-CNM), Consejo Superior de Investigaciones Científicas (CSIC), 08193 Bellaterra, Barcelona, Spain. [5]Technological Centre for Innovation in Communications (iDeTIC), University of Las Palmas de Gran Canaria, Las Palmas, Spain. [6]Centre for Marine Sciences of the Algarve (CCMAR), Universidade do Algarve, Campus de Gambelas, Faro, Portugal. [7]Aquaculture Research Group, Institute of Sustainable Aquaculture and Marine Ecosystems (IU-ECOAQUA), University of Las Palmas de Gran Canaria, Las Palmas, Spain.

*Correspondence:*
*Jaume Pérez-Sánchez*
*jaime.perez.sanchez@csic.es*

# ABSTRACT

Recent technological advances in microelectronics together with broad-band satellite communication and coverage make feasible remote observations of the behaviour and locomotion of aquatic and terrestrial animals in the wild. Biosensor devices are also increasingly used for a non-invasive understanding of basic biology during farming and experimental biology. The aim of this study was to design and validate customized and miniaturized tri-axial accelerometers for remote and non-invasive monitoring of farmed fish with re-programmable schedule protocols. The current package device (AE-FishBIT v.1s) is a non-wireless, stand-alone system that measures 14 mm length and 7.2 mm wide with a total mass of 600 mg, which allows monitoring animals from 30-35 g onwards. Validation experiments were performed in juveniles of gilthead sea bream and European sea bass, attaching the device to the fish operculum for monitoring physical activity by measurements of movement accelerations in x- and y-axes, while records of operculum breathing (z-axis) serve as a direct measurement of respiratory frequency. Data post-processing of exercised fish in swimming test-chambers revealed an exponential increase of fish accelerations with the increase of fish speed from 1 body-length to 4 body-lengths per second, while a close linear parallelism between oxygen consumption ($MO_2$) and operculum breathing was consistently found. Preliminary tests in free-swimming fish kept in rearing tanks also showed that biosensor data recording is able to detect changes in fish circadian fish activity. The usefulness of low computational load for data pre-processing with on-board algorithms was also verified from low to submaximal exercise, increasing this procedure (in combination with ultra-low-energy micro-programming) the autonomy of the system up to 6 h of continuous data recording with different programmable schedules. Visual observations regarding tissue damage, feeding behaviour and circulating levels of stress markers (cortisol, glucose, lactate) did not reveal at short term a negative impact of tagging. Although, reduced plasma levels of triglycerides revealed a transient inhibition of feed intake in small fish (sea bream 50-90 g, sea bass 100-200 g). This is the proof of the concept that miniaturized devices are suitable for non-invasive and reliable metabolic phenotyping of challenged fish to improve overall fish performance and welfare.

# INTRODUCTION

Biosensors based on the transduction of optical, electrical and mechanical signals are increasingly used to monitor a broad range of physiological and behavioural variables for both diagnostic and monitoring applications (Tamayo et al., 2013). The best example of optical devices is the finger pulse oximeter that detects $O_2$ content in the blood based on the way that the light passes through the finger. The electrocardiogram (ECG) sensors are also one of the most common medical tools used in modern medicine to assess the electric and muscular function of the heart. Likewise, the technology of accelerometers has the potential to revolutionize healthcare and sports medicine through low-cost and pervasive physiological devices for monitoring physical activity. This kind of mechanical transducer has been used long time ago in civil engineering, aviation and automotive industry for recording movement, velocity and acceleration magnitudes. However, with the advent of the MEMS (MicroElectroMechanical System) technology, high-precision and low-cost accelerometers are available for integration with microprocessors in consumer electronic components as different physical activity gadgets (activity trackers, fitness band and heart rate monitors, sport-watches, smart pedometers and wellness monitors, among others). This makes feasible the use of this type of devices to measure exercise routines and fitness regimens as well, but they are especially valuable when combined with wireless heart rate and ECG monitoring (Payne et al., 2014; Tamsin, 2015).

Recent technological advances in microelectronics, combining accelerometers and UHF RFID together with broad-band satellite communication and coverage, make feasible remote observations of the behaviour and locomotion of aquatic and terrestrial animals in the wild (Brown et al., 2013; Hussey et al., 2015; Wilson et al., 2015). Biosensor technology is also increasingly used for a non-invasive understanding of basic biology during farming and experimental biology of a range of variables that are directly or indirectly relevant for animal health and productivity (Ivanov et al., 2015; Neethirajan et al., 2017). Concretely, the Precision Fish Farming (PFF) concept based on the use of cameras, sonars, acoustic telemetry and environmental biosensors of chemical and biological safety risk factors seeks to harness the potential of technologically founded methods for improving fish production efficiency, product quality and disease prevention, reducing at the same time the environmental impact of farming operations (Føre et al., 2017; Saberioon et al., 2017; Enviguard EU project: www.enviguard.net). Recent studies also highlighted the use of micro-accelerometers to detect and identify different types of behavioural events, such as feeding and escape reactions (Broell et al., 2013). Furthermore, the concept of sentinel animals fitted with biosensors that feedback into farm-sensor network was initially developed for the dairy industry (Simbeye et al., 2014) and are currently being developed for oysters (Andrewartha et al., 2015) and farmed fish (Staaks et al., 2015), existing abundant literature on measures of body temperature, animal orientation and depth in both wild and farmed fish (e.g. Clark et al., 2008; Payne et al., 2014; Metcalfe et al., 2016, among others).

Acceleration data-loggers alone or in combination with pressure, temperature and heart-rate biosensors have also been used for tracing movement and estimating activity-specific energy expenditure or feeding behaviour in a number of fish species, including juvenile hammerhead sharks (*Sphyrna lewini*, Gleiss et al., 2010), sockeye salmon (*Onchorhynchus nerka*, Clark et al, 2010), European sea bass (*Dicentrarchus labrax*, Wright et al., 2014), Atlantic salmon (*Salmo salar*, Kolarevic et al., 2016) and red-spotted groupers (*Epinephelus akaara*, Horie et al., 2017). These sensors operate as acoustic transmitter tags that contain a tri-axial accelerometer, which registers gravity forces and acceleration in the x-, y- and z-directions.

Currently, the smallest market size for these tags is 7 mm diameter and 20 mm length with a weight of 2.6 g in air and a typical battery life of 1-3 months, depending on measurement periods and transmission intervals. However, as pointed out by Kolarevic et al. (2016), the interference between transmitted signals limits the use of large number of acceleration tags in a single rearing unit or in two neighbouring units. Otherwise, in the aquatic environment, the use of low radiofrequency transmission is limited to 1-2 m of maximum communication range without repeaters (Palmeiro et al., 2011). Thus, with the double aim to produce small biosensors and to combine measures of physical activity and energy demand, we prompted us to design and produce within the AQUAEXCEL$^{2020}$ EU project ultra-low power stand-alone devices to be implanted on fish operculum for measurements of physical activity and respiratory frequency. To the best of our knowledge, there is not in the market any device designed to provide simultaneously these two types of measures. The functional testing of the first prototype (AE-FishBIT v.1s) was conducted as the proof of concept in European sea bass and gilthead sea bream (*Sparus aurata)* as the two most significant farmed fish of the Mediterranean aquaculture (FAO, 2018).

## MATERIALS AND METHODS

### Hardware architecture

The proposed device is a programmable and reconfigurable tri-axial accelerometer for measuring accelerations in x-, y- and z-axes in the range of ± 8 *g* and sampling frequencies up to 800 Hz. The basic components disposed on a 0.8 mm 4-layer printed circuit board (pcb) are: 1) one accelerometer (MMA8451Q; NXP Semiconductors Research, Eindhoven, The Netherlands), 2) one high performance Li-ion battery (UMAC0401; Murata Electronics, Kyoto, Japan) of low charge time (120 sec), 3) one microprocessor (KL17; NXP Semiconductors Research) with 256 kbyte flash memory and 32 kbyte RAM memory and 4) one RFID tagging device (nano-transponder ID-100A-1.25; TROVAN, Madrid, Spain) for rapid identification. System integration included water-proof packaging (tested up to 6 bars of pressure in air and seawater environments) with isolated connector pins. The complete package size of the AE-FishBIT v.1s measures 14 mm length and 7.2 mm wide with a total mass of 600 mg in air (**Figure 1**).

### Biosensor location and attachment procedure

The operculum was chosen as the target location for the biosensor as it allows monitoring physical activity by measurements of accelerations in x- and y-axes, while records of operculum breathing (z-axis) serve as a direct measurement of respiratory frequency. The device attachment was accomplished using small and light laboratory tags for identification of experimentation animals (RapID tags from RapID Lab, Inc, San Francisco, CA, USA) that were rapidly pierced to opercula of anaesthetized fish with 100 μg/mL 3-aminobenzoic acid ethyl ester (MS-222, Sigma, Saint Louis, MO, USA). In a second step, a rigid 3D-printed pocket was fixed with an innocuous and quick drying aquarium adhesive in the water (cyanoacrylate) to the exterior side of the RapID tag. Such approach allowed easy application and removal of the device in the pocket (**Figure 2**).

**Data acquisition and software processing**

The software processing aims at estimating the respiratory frequency and an index of the fish activity. The first estimation is carried out with the signal of the z-axis of the accelerometer, and the second estimation is obtained from the x- and y-axis signals of the accelerometer.

The recorded z-axis accelerations generate positive and negative signals, related to operculum opening and closing, respectively. Other movements superimposed to this z-axis signal were angular accelerations that inform of fish trajectory or side head movement, though the dominant periodic component is always the operculum movement. To calculate the respiratory frequency, the z-axis signal of the accelerometer is band pass filtered between 0.5 and 8 Hz to reduce the noise and highlight the periodic properties of the signal. The numbers of maxima or minima in the recorded signals are registered at a sampling rate of 100 samples per second. The measurements were expected below 5 Hz, and the z-axis signal was bandpass filtered between 0.5 and 8 Hz to highlight the periodic properties of the signal. Then, the signal was derived and the number of crosses through zero was divided by two to obtain the signal period.

To alleviate the drawback of the noise in the accelerometer signals and to obtain a more accurate estimation of the number of peaks, they are estimated in $N$ consecutive frames of $T$ seconds. The final number of peaks $N_p$ is calculated as the 25% percentile of the $N$ estimations. With this quartile, the less noisy frames were selected, avoiding complex noising signal processing algorithms that would be very high-energy demanding for the device microprocessor. Hence, the respiratory frequency is issued every $N \cdot T$ seconds and it is calculated as:

$$F_{resp} = N_p/T$$

Values of $N, T$ and $f_s$ were heuristically chosen to assure a good statistical representation. The value of $f_s$ was established as 100 Hz. A value of $T_s = 10$ seconds was chosen as frame length. This value supposes 5 peaks if the respiratory frequency is 2 Hz. $N$ was established equal to 12, which allowed a reasonable estimation of the 25% percentile. This values means that the respiratory frequency is calculated every 2 minutes (120 seconds).

The physical activity index described the intensity of the fish movement in the x- and y-axes, and it was based on the energy of the jerk. The jerk is the derivative of acceleration and its magnitude is completely orientation-independent (Hamäläinen et al., 2011). Similarly to the respiratory rate, the energy of the jerk was estimated on $N$ consecutive frames of $T$ seconds and the percentile 25% provided as activity index measure. The values of $N$, $T$ and $f_s$ are the same than in the respiratory frequency algorithm. The accelerations in the x- and y-axis, named $a_x(n)$ and $a_y(n)$ respectively, were derived as:

$$d_x(n) = a_x(n) - a_x(n-1)$$
$$d_y(n) = a_y(n) - a_y(n-1)$$

The standard deviation of $d_x(n)$ and $d_y(n)$ were calculated:

$$\sigma_x = \sqrt{\frac{\sum_{n=1}^{T \cdot f_s}(d_x(n)-\mu_x)^2}{T \cdot f_s}} \quad \text{and} \quad \sigma_y = \sqrt{\frac{\sum_{i=1}^{T \cdot f_s}(d_y(n)-\mu_y)^2}{T \cdot f_s}}$$

where $\mu_x$ and $\mu_y$ are the average of $d_x(n)$ and $d_y(n)$ respectively. The energy of the jerk is then obtained as:

$$E_{jerk} = \sqrt{\sigma_x^2(n) + \sigma_y^2(n)}$$

The physical activity index is the percentile 25% of the energies of the jerk estimated on $N$ consecutive frames. A summary of the procedure for the calculation of respiratory frequency and physical activity index is shown in **Figure 3**.

Algorithms for direct on-board calculation of respiratory frequency and physical activity index by the biosensor were programmed through the FRMD-KL25Z algorithm platform and uploaded to the device. This step involved some changes on the algorithm to estimate the physical activity index due to the limited computational power of the biosensor. For this purpose, the standard deviation of $d_x(n)$ and $d_y(n)$ was approximated by:

$$\sigma_x = \frac{\sum_{n=1}^{T \cdot f_s}|d_x(n)-\mu_x|}{T \cdot f_s} \quad \text{and} \quad \sigma_x = \frac{\sum_{n=1}^{T \cdot f_s}|d_y(n)-\mu_y|}{T \cdot f_s}$$

The energy of the jerk was estimated as:
$$E_{jerk} = \sigma_x + \sigma_y.$$

The value of $T$ were changed from 10 second to 10.24 seconds. It means a frame of 1024 data, which is a power of 2, more convenient for the biosensor programming. Data post-processing in each measured interval maximized the data storage in the memory of data-logger. The platform also allowed firmware updates and data downloading (raw data or post-processed data) to a computer with a serial wire debug cable.

**Functional testing**

The initial test of the device was conducted in sea bream and sea bass juvenile fish (30-150 g body weight), exercised in an intermittent-closed swim tunnel respirometer of 10 L water volume (Loligo® Systems, Viborg, Denmark). The swim tunnel was submerged into a water bath that served as a water reservoir for flushing the respirometer after each closed respirometer run (flush pump: Eheim 1048, 10 L/min; Deizisau, Germany). To ensure constant high water quality, the water bath was connected by a second flush pump (Eheim 1250, 20 L/min; Deizisau, Germany) to a 100 L-reservoir tank coupled to a re-circulatory system equipped with physical and biological filters and a programmable temperature system fixed at 24-25 ºC. Sea water (80-95% $O_2$ saturation) flowed back into the re-circulatory system by means of gravity, remaining unionized ammonia, nitrites and nitrates almost undetectable along the whole experiment. A thruster within the respirometer was used to generate a swimming current. Water velocities were calibrated with a hand-held digital flow meter that was ordered by the controller Movitrac® LTE 0.37kW/0.5HP (SEW Eurodrive, Normanton, United Kingdom). Respirometry runs and chamber flushing were automatically controlled with the DAQ-M instrument (Loligo® Systems) connected to a PC equipped with AutoResp™ software (Loligo® Systems). Water temperature and $O_2$ saturation within the respirometer was measured using a Witrox 1 single channel oxygen meter (Loligo® Systems, Viborg, Denmark), equipped with a needle-type fibre optical micro-sensor (NTH, PreSens-Precision Sensing GmbH, Regensburg, Germany) and a temperature probe suspended into the water current within the respirometer.

For testing procedures, anaesthetized fish with the AE-FishBIT v.1s prototype were transferred into the swim tunnel, and allowed to recover and acclimate at a swimming speed of 0.5-1.0 body-lengths per second (BL/s) until their measurements of $O_2$ consumption rates ($MO_2$) reached a constant low plateau. This typically happened after 30-45 min with $MO_2$ around 220-240 $mgO_2$/kg/h (Martos-Sitcha et al., 2018b). After this acclimation period, water velocity was increased in 0.5 BL/s steps, and specimens were submitted to controlled speeds between 1-4 BL/s during the first testing approaches, or from 1 BL/s until exhaustion in the final on-board validation of data recording. Each swimming interval at a given velocity lasted 5 min, consisting in "flush-wait-measurement" cycles (60 sec flush interval to exchange the respirometer water = "flush"; 30 sec mixing phase in closed mode = "wait"; and a 210 sec $MO_2$ measuring period in closed mode). During the measurement interval, $O_2$ saturation of the swim tunnel water was recorded every second. $MO_2$ was automatically calculated by the AutoRespTM software from linear decreases ($r^2$ = 0.98-1.0) in chamber $O_2$ saturation using the appropriate constants for $O_2$ solubility in seawater (salinity, temperature and barometric pressure). For each fish, three data sets of 2 min raw data were acquired by the biosensor at different speed steps. After the swim tunnel test, each fish was recaptured and again anaesthetized to plug-out and recuperate the biosensor for data download and post-processing on a PC or directly by on-board algorithms.

For tests conducted with free-swimming fish in 500-L tanks, the biosensor devices were programmed to acquire 2 min raw data sets at three discrete times one day after device implantation (10:00 a.m., 2:00 p.m. and 6:00 p.m.). This schedule covers a wide range of possible circadian variations in activity and respiration indexes. Additional tests with active feeding fish reared at 22-24 ºC were conducted with dummy devices (same size and weight than the functional prototype) to check how blood markers of stress and welfare were affected by the prototype implantation in both sea bass (100-200 g) and sea bream fish of two different class of size (50-90 g; 300-500 g). After one week, tagged and non-tagged fish (n = 5-7 fish per group were anesthetized and blood was quickly taken from caudal vessels with heparinized syringes. A blood aliquot was centrifuged at 3,000 x *g* for 20 min at 4ºC, and the plasma was stored at -80 ºC until the subsequent biochemical assays. Plasma glucose levels were measured by the glucose oxidase method (ThermoFisher Scientific, Waltham, Massachusetts, USA) adapted to 96-well microplates. Blood lactate was measured in deproteinized samples (perchloric acid 8%) using an enzymatic method based on the use of lactate dehydrogenase (Instruchemie, Delfzijl, The Netherlands). Plasma triglycerides (TG) were determined using lipase/glycerol kinase/glycerol-3-phosphate oxidase reagent (ThermoFisher Scientific). Plasma cortisol levels were analysed using an EIA kit (Kit RE52061 m IBL, International GmbH, Germany). The limit of detection of the assay was 2.46 ng/mL with intra- and inter-assay coefficients of variation lower that 3% and 5%, respectively.

No mortality was observed during all the experimental period and all the described procedures were approved by the Ethics and Animal Welfare Committee of Institute of Aquaculture Torre de la Sal and carried out according to the National (Royal Decree RD53/2013) and the current EU legislation (2010/63/EU) on the handling of experimental fish.

**Statistical analysis**

Daily variation of respiratory frequency and physical activity index in free-swimming sea bream and sea bass was analysed by one-way ANOVA. Blood parameters of tagged and non-tagged groups were compared by means of Student's t-test. Significance levels were set at P < 0.05. Analyses were performed using the SigmaPlot Version 13 for Windows (Systat Software Inc., Chicago, IL, USA). Multivariate partial least-squares discriminant analysis (PLS-DA) of data on respiratory frequency and physical activity index of exercised sea bream in the 1-6 BL/s water speed range was conducted with the EZ-info software (Umetrics, Sweden). The quality of the PLS-DA model was evaluated by $R^2Y$ and $Q^2$ parameters, which indicated the fitness and prediction ability, respectively.

## RESULTS

Raw data generated by the tri-axial accelerometer was stored in the memory of the device in the initial tests, and they were processed after download in a computer for calculations of the respiratory frequency and the physical activity index. Using this methodology, the autonomy of the device is limited to the amount of memory, which is able to store a total of 6 min of raw data measured at a rate of 100 samplings per second. It was established that 2 min is a reliable time window for statistical calculations, and the device was then programmed to acquire three sets of 2 min of raw data for a given experimental schedule. Software development also offered the possibility to process on-board recorded data by means of the proposed mathematical approximations to ease microprocessor function. Such approach is not a high memory consuming process, and the autonomy of the system was consequently increased up to 6 hours of continuous data recording, allowing longer schedules as long as the battery of the device is operative. Initials tests were conducted for the assessment of the functional significance of calculated outputs from raw data post-processed on PC. On a second step, the suitability of on-board calculations (compared with those derived from raw data extraction) was assessed as detailed below.

Initials tests in swim tunnel respirometer showed for both sea bream (**Figure 4A**) and sea bass (**Figure 4B**) a linear increase of $MO_2$ in the 1-4 BL/s speed range. The calculated respiratory frequency by data post-processing on PC paralleled $MO_2$. At the same swimming speed, $MO_2$ and the calculated respiratory frequency were consistently higher in sea bream than in sea bass. However, for a given $MO_2$, the respiratory frequency was similar for both species as shown by the correlation plot of these two variables when all data for both species were put together (**Figure 4C**). Regarding physical activity index, it increased exponentially rather than linearly with the increase of swimming speed, as the total jerk magnitude only reflects fish-related accelerations (**Figure 5**). This pattern was observed both in sea bream and sea bass, though it becomes more evident in sea bream at swimming speeds over 3 BL/s.

Values registered in free-swimming fish were in the measurable range by the PC-tested algorithms in the swim tunnel. In both fish species, the calculated respiratory frequency in holding tanks did not show significant daily variations from 10:00 a.m. to 6:00 p.m, ranging the respiratory frequency between 2.4-2.3 breath/s in sea bream and 2.2-1.8 breath/s in sea bass (**Figure 6A**). For the measurements of physical activity, the range of variation was also higher in sea bass, being statistically significant the decrease (relative units) of physical activity from 0.15 to 0.03 ($P < 0.05$). The observed decrease in sea bream varied from 0.15 to 0.09 ($P = 0.12$) (**Figure 6B**).

For validation of on-board algorithms, sea bream juveniles of 80-100 g were exercised over 1-6 BL/s until exhaustion (fish were considered exhausted when they rested at the back grid for at least 5 sec) in the swim tunnel, and data processing of both respiratory frequency and physical activity index evidenced close linear correlations near to 1 between on-board and PC-calculated values (**Figure 7**). The same data were analysed to underline the different dynamics of analysed parameters from low to submaximal exercise, and it was observed that the respirometer measurement of $MO_2$ increased linearly up to a swim speed of 4.5 BL/s, paralleling the increase in gill breathing (on-board algorithms) with a maximum respiratory frequency (MRF) close to 4 BL/s (**Figure 8A, 8B**). This short delay in the achievement of the maximum metabolic rate (MMR), defined as the maximum $O_2$ consumption in exercised fish, might be due, at least in part, to the instantaneous nature of biosensor measurements (gill breathing) compared to the buffered changes in the measurements of $O_2$ concentrations in the 10-L chamber of respirometer. In any case, these two types of measures of respiration decreased progressively and markedly with the increased contribution of anaerobic metabolism near to submaximal exercise. Likewise, measurements of physical activity processed by on-board algorithms evidenced a maximum activity at 5 BL/s. This was preceded by a slight decrease of slope at 4.5 BL/s that becomes clearly negative with the enhancement of the unsustainable anaerobic metabolism close to submaximal exercise (**Figure 8C**). Any of the analysed variables is informative enough to ascertain the aerobic/anaerobic scope when considered individually. However, given their different dynamic patterns in response to exercise, multivariate analyses with on-board processed data resulted in a fairly good differentiation of aerobic/anaerobic fish condition along the first component, with a 59% of total variance explained ($R^2$) and 57% of total variance ($Q^2$) predicted by the two components of the discriminant model (**Figure 9**).

The impact of biosensor attachment in fish physiology was assessed by comparisons of circulating levels of markers of stress and welfare in tagged and non-tagged free-swimming fish one week after implantation of dummy devices. No differences were found in circulating levels of cortisol (**Figure 10A**), glucose (**Figure 10B**) or lactate (**Figure 10C**) in 100-200 g sea bass nor sea bream in the two class of size analysed. By contrast, TG levels in 50-90 g sea bream and 100-200 sea bass evidenced a significant decline ($P < 0.01$) in tagged fish in comparison with non-tagged ones (**Figure 10D**). This disturbance was not detected in larger fish (> 300 g sea bream).

**DISCUSSION**

The present study is the proof of concept of the work conducted within the AQUAEXCEL[2020] EU project for the design, programming and testing of a miniaturized device (AE-FishBIT v.1s) for individual fish phenotyping of metabolic condition and welfare. Different types of biosensors available as research tools or commercial products were initially considered (Tamayo et al., 2013; Bandodkar and Wang, 2014; Tamsin, 2015), but given the methodological and cost limitations underwater, it was decided that the use of mechanical biosensors is more feasible than other possible solutions based on electric and/or optical sensors. The final developed device is a miniaturized tri-axial accelerometer that is able to register (two in one) physical activity and opercula breathing as a direct measure of respiratory frequency. Due to the constraints in size, weight, battery consumption and signal transmission in aquatic environments (Lloret et al., 2012; Climent et al., 2014), we decided to prioritize the production of sensor devices working in stand-alone mode (no wireless data transmission). The system also includes a tag RFID system for an easier operability and data processing during fish data recording from resting to moderate or very active behaviour under aerobic and/or anaerobic conditions.

When the system was tested in exercised juveniles, close lineal correlations near to 1 were found between $O_2$ consumption and respiratory frequency in both sea bream and sea bass. The close associations between the Loligo® Systems respirometer and AE-FishBIT biosensor data outputs also applies to measures of physical activity, though in this case the jerk of accelerations increased exponentially rather that linearly with the increase of swimming speed (**Figures 4, 5**). Indeed, the jerk magnitude is independent of orientation and it only reflects accelerations, which are theoretically zero at constant speed. Thus, the vigorous and irregular tail movements of fish at high speed resulted in a non-linear increase of the jerk of accelerations. This activity feature was more evident in the case of sea bream, which probably reflects a lower capability for fast swimming in comparison to sea bass with the spindle-shaped body characteristic of an active predator (Spitz et al., 2013), as indicates its common name in France "le loup de mer". Likewise, for a given $MO_2$, measurements of jerk accelerations were higher in free-swimming fish than in forced exercised fish (swim tunnel), where a more static position limits changes of trajectory and accelerations (**Figures 4, 5, 6**). From tests conducted across the light-phase in free-swimming animals, it is also conclusive that the range of variation becomes higher for measurements of physical activity rather than respiratory frequency, especially in the case of sea bass (**Figure 6**). However, this issue needs to be corroborated with a more continuous circadian data recording with the advent of new AE-FishBIT versions.

The usefulness of monitoring behaviour remains yet to be fully exploited in aquaculture practice, though it is well known that swimming performance provides a complete measure of animal fitness and perhaps stress behaviour in fish kept under specific environmental conditions (Nelson, 1989; Plaut, 2001; Remen et al., 2016). Thus, measurements of $MO_2$, and secondly respiratory frequency, can be considered a good measure of the energy expended by fish to integrate a wide-range of physiological processes influenced by both endogenous (e.g., size, age, fish strain) and exogenous (light, temperature, $O_2$ concentration, time of day, etc) factors. In our experimental conditions, $MO_2$ increased in sea bream and sea bass from 230-270 to 370-400 $mgO_2/kg/h$ with the increase of swimming speed from 1 to 4 BL/s (**Figure 4A**). These reported values are in line with those found in previous studies of gilthead sea bream (Martos-Sitcha et al., 2018b), European sea bass (Claireaux et al., 2006) or other farmed fish, such as meagre (*Argyrosomus regius*) (Peixoto et al., 2016), challenged at water temperatures close to those set in our experimental approach. Thus, given the close parallelism between measures of $MO_2$ and respiratory frequency, we can conclude that the AE-FishBIT biosensor provides reliable results of $O_2$ consumption in a wide-physiological range not only when the energy requirements do not exceed the $O_2$ demand for a sustainable aerobic metabolism, but also during the anaerobic phase that is largely increased at submaximal exercise (Ejbye-Ernst et al., 2016). This was evidenced by additional swimming tests (**Figures 7, 8, 9**), prolonged until fish exhaustion at 6 BL/s, which also served to corroborate the suitability of on-board algorithms of respiratory frequency and physical activity through aerobic and anaerobic conditions (see below).

The weight of the first AE-FishBIT prototype is 600 mg in air with an estimated 50 % of buoyancy. These features make it very convenient for its use in fish of 30-35 g onwards, according to the empirical "2 % tag/body mass" rule for implanted devices (Winter, 1996; Jepsen et al., 2004). Recent works claim that this rule can be even extended for some fish species up to 7 % of body mass without detrimental effects on performance or survival (Smircich and Kelly, 2014; Makiguchi and Kojima, 2017). In the present study, we assessed the influence of biosensor implantation on fish welfare by measurements of blood biochemical parameters in both sea bream and sea bass, and no differences were found between tagged and non-tagged fish in classical stress markers (cortisol, glucose, lactate) one week after device attachment regardless of body weight (50- >300 g). During this time, the observed feeding behaviour was also considered to be quite similar regardless of device implantation. Although, the lowering effect on plasma TG levels of device attachment in 100-200 g sea bass and 50-90 g sea bream would be indicative of, at least, a transient impairment of feed intake. This finding was not evidenced in the largest sea bream (> 300 g), being this fact indicative of the importance to define the critical fish size to minimize the impact of tag burden in welfare and behaviour in different aquaculture scenarios (**Figure 10**). In parallel, important research efforts continue to be made to reduce the dimensions and weight of upcoming AE-FishBIT prototypes, keeping or even increasing the functional features of the device. At short-medium term, the main envisaged hardware strategies for this miniaturization will consider the use of: i) flexible kapton instead of the current pcb rigid substrate, ii) lighter and smaller long-life batteries and iii) improved circuit packaging architecture based on Flip Chip (Lau, 2016) or embedded wafer level ball grid array (eWLB) technologies (Meyer et al., 2008). These solutions are based on currently available commercial components, but as a part of semi-industrialization packaging and production, we cannot exclude the design of custom integrated circuits as the most ambitious and technologically challenging procedure for the final product miniaturization.

Data processing and software improvements with on-board algorithms are also key steps on the prototype development due to its major impact on data recording autonomy, which makes possible different short- and long-time schedules, adjusted for instance to 2 days (2 min recording each 15 min), one week (2 min recording each 60 min) or three weeks (2 min recording each 180 min). This relatively high autonomy also requires a low computational load, which did not compromise the usefulness of the microprocessor mathematical approximations. This was evidenced by close linear correlations of PC and on-board biosensor outputs in sea bream juveniles, exercised from low to submaximal exercise (**Figure 7**). This long coverage of exercise activity highlighted that the system remains highly informative of metabolic condition with the shift of aerobic to unsustainable anaerobic metabolism, which reduces the efficiency of ATP production and promotes the accumulation of deleterious by-products (eg. $H^+$, lactate) (Richards, 2009; Seibel, 2011). Indeed, aerobic locomotor activity is powered by red oxidative muscle fibers, but when approaching their maximum power capacity, a gait transition to anaerobic fueling is assisted by the activation of fast white muscle fibers that results in fatigue and depletion of muscle glycogen depots (Kieffer, 2000; Sänger and Stoiber, 2001). Some degree of activation of anaerobic metabolism occurs before reaching MMR, as it has been evidenced in different experimental models (Burgetz et al., 1998; Lee et al., 2003; Hinch et al., 2006; Teulier et al., 2013), including sea bream and guppy (*Poecilia reticulata*) (Ejbye-Eernst et al., 2016). However, as pointed out before, the closely related MRF and MMR will mark the start of a burst-assisted swimming, resulting in a change of metabolic scope and swimming energy partitioning (Peake and Farrell, 2004; Peake, 2008; Marras et al., 2013; Svendsen et al., 2015). Importantly, this energy transitioning can be monitored by our prototype device, because discriminant multivariate analysis distinguished two main groups of fish behavior when the achieved MRF-MMR was taken as main criteria of aerobic/anaerobic classification (**Figure 9**). This complex trade-off is mediated by $O_2$ sensors (Norin and Clark, 2016), being defined the oxygen limiting saturation (LOS) as the $O_2$ threshold level that is no longer sufficient to maintain $MO_2$ at a given temperature and voluntary swimming activity (Remen et al., 2013, 2015). Experimental evidence also indicates that acute hypoxia drives to the re-adjustment of mitochondrial machinery at transcriptional level to cope with a decreased basal metabolic rate, consistent with a low risk of oxidative stress, diminished aerobic energy production and higher $O_2$-carrying capacity in blood cells (Martos-Sitcha et al., 2017). Different types of adaptive responses also occur under moderate hypoxia, which reflects the tissue-specific responsiveness of liver, heart, skeletal muscle and blood cells, according to their metabolic capabilities and $O_2$ availability (Martos-Sitcha et al., 2019). Fish transcriptomic meta-analysis in response to a vast array of challenged conditions highlights the key role of mitochondria in front of different cellular stresses, including hypoxia, hypercortisolism and malnutrition (Calduch-Giner et al., 2014). In particular, sea bream juveniles exposed to thermal stress or multiple sensory perception stressors (shaking, sound, light flashes, water flow reversal) evidence adaptive responses of glycolytic pathways and mitochondrial respiratory chain (Bermejo-Nogales et al., 2014). How these adjustments at cellular level can be correlated with the monitored AE-FishBIT parameters is the upcoming envisaged challenge in a scenario of global change with increases of temperature, ocean acidification and reduced $O_2$ concentrations (Gruber, 2011).

Another important challenge is the improvement of the biosensor attachment procedure to the operculum, which was chosen as the target location to make feasible the simultaneous measurements of respiration- and activity-related parameters. External tagging procedures in fish are usually related to telemetry devices in large fish, with application near the dorsal or anal fins (Jepsen et al., 2015). At the present stage, the RapID tag & 3D pocket approach has served for the short-term validation experiences, but this system procedure is not operative for more than 1-2 weeks due to the appearance of macroscopic signals of damage and necrosis caused by loosening at the piercing location. This problem has been currently solved with the use of corrosion-resistant self-piercing fish tags (National Band and Tag Co., Newport, KY, USA) as the attaching biosensor support. In our hands, the attachment implantation procedure takes less than 1 min, but automatization procedures like those used for massive fish vaccination (Skala Maskon, Stjørdal, Norway) need to be implemented for a more practical and routine use of biosensors in fish farming. In any case, specific attachment procedures need to be validated for each fish species, size and physiological condition in different aquaculture scenarios through production cycles.

Further improvements envisaged in the AE-FishBIT device include a more compact design (**Figure 11**) with the reallocation of the connection pins for charge and data transmission in the bottom side of pcb. The electronic components will be located in the top layer. The enclosure of the device is a protective packaging that completely covers the edge of the printed circuit, which will improve the insulation of the biosensor components. Additionally, the new printed circuit is manufactured in multilayer technology without through holes. The combination of the compact pcb and the protective enclosure results in a fully sealed biosensor with a minimum increase of the total weight (around 900 mg). From the electrical point of view, the design can stand water contact with the connector pins. The supply voltage of the battery is protected by a diode that prevents its discharge. During the experiments, data transmission signals are always in the high impedance state, which prevents currents leakage. The intended prototype (AE-FishBIT v.2) has been tested underwater for more than 24 hours without any additional protection in the electrical contacts. For longer experimental periods, the electrical contacts will be further protected with a removable waterproof lid fixed with strong adhesive. The removable cap will be adapted to be combined with the best attachment procedure for each fish species and experiment duration.

**Concluding remarks**

A miniaturized device to register at the same time fish operculum breathing (respiratory frequency) and physical activity has been designed, produced and tested in sea bream and sea bass as the proof of concept of the prototype. The basic operating mode is stand-alone with an autonomy of 6 h of continuous data recording with different programmable schedules (e.g. 2 min each 15 m for 2 days; 2 min each 60 min for 1 week; 2 min each 180 min for 3 weeks). Validity and functional significance of tri-axial accelerations has been assessed under forced and voluntary exercise and further work is underway to improve semi-industrial packaging and production. From a functional point of view, data analysis depicts the potential use of the proposed prototype for assessing the gait transition of metabolic scope and energy partitioning. The expected impact is the improvement of metabolic phenotyping in a poor invasive manner for the implementation of fish management and selective breeding. The prototype can also contribute to establish more strict and reliable welfare standards, and a better perception of quality controls in aquaculture production.

The prototype is protected by a registered patent (76763/P7259).

**FIGURE CAPTIONS**

**Figure 1. (A)** Electrical probing of pcb-mounted devices. **(B)** Photograph of AE-FishBIT v.1s before encapsulation. **(C)** Encapsulated AE-FishBIT v.1s in underwater operation. Flashing led light indicates the end of data acquisition program for user reference.

**Figure 2.** Biosensor attachment to sea bream operculum. **(A)** RapID tag is attached to fish operculum with the tag piercing tool. The external side of the tag **(B)** allows fixation with adhesive of a 3D-printed pocket. **(C)** Internal side of attached sea bream operculum. **(D)** Sea bream with a biosensor attachment with the RapiID tag & 3D pocket procedure.

**Figure 3.** Procedure diagram for the calculation of respiratory frequency and physical activity index.

**Figure 4. (A)** Measures of $MO_2$ (mg$O_2$/kg/h; grey circles) and respiratory frequency (breaths/s; black triangles) of sea bream with increasing speed in the swim tunnel respirometer. Each point represents mean±SEM of 8 measures. **(B)** Measures of $MO_2$ (mg$O_2$/kg/h; grey circles) and respiratory frequency (breaths/s; red triangles) of sea bass with increasing speed in the swim tunnel respirometer. Each point represents mean±SEM of 8 measures. **(C)** Correlation plot of $MO_2$ and respiratory frequency values for each sea bream (black circles) and sea bass (red circles) individual.

**Figure 5.** Output of physical activity index with increasing speed in the swim tunnel respirometer for sea bream (black circles) and sea bass (red circles). Each point represents mean±SEM of 8 measures.

**Figure 6. (A)** Daily variation of respiratory frequency values in free-swimming sea bream (white bars) and sea bass (black bars) in 500-L tanks. Each bar represents mean±SEM of 6 measures. **(B)** Daily variation of physical activity values in free-swimming sea bream (white bars) and sea bass (black bars) in 500-L tanks. Each bar represents mean±SEM of 6 measures. Different superscript letters indicate significant differences ($P < 0.05$; one-way ANOVA).

**Figure 7.** Validation of on-board algorithms from low to submaximal exercise. Values are mean of six sea bream juveniles. **(A)** Correlation plot for a given swimming speed between respiratory frequency values calculated for 2 min raw data post-processed on a PC (x-axis) or on-board (y-axis). **(B)** Correlation plot for a given swimming speed between physical activity index values calculated for 2 min raw data post-processed on a PC (x-axis) or on-board (y-axis)

**Figure 8.** Respirometer and on-board biosensor output from low to submaximal exercise. Values are the mean±SEM of six sea bream juveniles **(A)** Respirometer measurements of $MO_2$ (mgO$_2$/kg/h) with increasing swimming speed. Maximum metabolic rate (MMR) is marked. **(B)** Respiratory frequency (breaths/s) with increasing swimming speed. The maximum respiratory frequency (MRF) is marked. **(C)** Physical activity index with increasing swimming speed.

**Figure 9**. PLS-DA score plot of on-board calculated parameters at different swimming speeds before (black squares) and after (red squares) the achievement of the MMR in sea bream juveniles from low to submaximal exercise.

**Figure 10.** Plasma levels of **(A)** cortisol, **(B)** glucose, **(C)** lactate and **(D)** triglycerides as a percentage of non-tagged control values. Each bar represents mean±SEM of n = 5-7 for each fish species and class of size. Asterisks indicate statistically significant differences with control (P< 0.01, Student's t-test).

**Figure 11.** Schematic 3D view of the intended AE-FishBIT v.2 prototype before (**A**) and after packaging (**B**).

**AUTHOR CONTRIBUTIONS**

JPS coordinated the different research teams; JS, DRV and JAMN assembled the device components and implemented software programming; CCD and MAF established mathematical parameters for data outputs and on-board algorithm approximations; FJB, EC, AV and ML worked on device packaging and insulation procedures; JPS, JAMS and JMA conducted functional tests; JAMS, JACG and JPS wrote the manuscript; all authors edited the manuscript; all authors read and approved the final manuscript.

**CONFLICT OF INTEREST STATEMENT**

The authors declare that the research was conducted in the absence of any commercial or financial relationships that could be construed as a potential conflict of interest.

**ACKNOWLEDGMENTS**

We acknowledge support of the publication fee by the CSIC Open Access Publication Support Initiative through its Unit of Information Resources for Research (URICI).

**FUNDING**

This project has received funding from the European Union's Horizon 2020 research and innovation programme under grant agreement No.652831 (AQUAEXCEL[2020], Aquaculture infrastructures for excellence in European fish research towards 2020). This publication reflects the views only of the authors, and the European Commission cannot be held responsible for any use which may be made of the information contained therein. Additional funding was obtained from project SURF (TEC2014-60527-C2-1-R) of the Spanish Ministry of Economy and Competitiveness and the European Regional Development Fund (FEDER).

# REFERENCES

Andrewartha, S. J., Elliott, N. G., McCulloch, J. W., and Frappell, P. B. (2016). Aquaculture sentinels: smart-farming with biosensor equipped stock. *Journal of Aquaculture Research and Development* 7: 393. doi:10.4172/2155-9546.1000393

Bandodkar, A. J., and Wang, J. (2014). Non-invasive wearable electrochemical sensors: a review. *Trends in Biotechnology* 32, 363-371. doi: 10.106/j.tibtech.2014.04.005

Bermejo-Nogales, A., Nederlof, M., Benedito-Palos, L., Ballester-Lozano, G. F., Folkedal, O., Olsen, R. E., et al. (2014). Metabolic and transcriptional responses of gilthead sea bream (*Sparus aurata* L.) to environmental stress: New insights in fish mitochondrial phenotyping. *General and Comparative Endocrinology* 205:305-315. doi: 10.1016/j.ygcen.2014.04.016

Broell, F., Noda, T., Wright, S., Domenici, P., Steffensen, J. F., Auclair, J. P., et al. (2013). Accelerometer tags: detecting and identifying activities in fish and the effect of sampling frequency. *Journal of Experimental Biology* 216, 1255-1264. doi: 10.1242/jeb.077396

Brown, B. D., Kays, R., Wikelski, M., Wilson, R., and Klimley, A. P. (2013). Observing the unwatchable through acceleration logging of animal behaviour. *Animal Biotelemetry* 1:20. doi: 10.1186/2050-3385-1-20

Burgetz, I. J., Rojas-Vargas, A., Hinch, S. G., and Randall, D. J. (1998). Initial recruitment of anaerobic metabolism during sub-maximal swimming in rainbow trout (*Oncorhynchus mykiss*). *Journal of Experimental Biology* 201, 2711-2721.

Claireaux, G., Couturier, C., and Groison, A. L. (2006). Effect of temperature on maximum swimming speed and cost of transport in juvenile European sea bass (*Dicentrarchus labrax*). *Journal of Experimental Biology* 209, 3420-3428. doi: 10.1242/jeb.02346

Clark, T. D., Taylor, B. D., Seymour, R. S., Ellis, D., Buchanan, J., Fitzgibbon, Q. P., et al. (2008). Moving with the beat: heart rate and visceral temperature of free-swimming and feeding bluefin tuna. *Proceedings of the Royal Society of London B* 275, 2841-2850. doi: 10.1098/rspb.2008.0743

Clark, T. D., Sandblom, E., Hinch, S. G., Patterson, D. A., Frappell, P. B., and Farrell, A. P. (2010). Simultaneous biologging of heart rate and acceleration, and their relationships with energy expenditure in free-swimming sockeye salmon (*Oncorhynchus nerka*). *Journal of Comparative Physiology B* 180, 673-684. doi: 10.1007/s00360-009-0442-5

Climent, S., Sanchez, A., Capella, J., Meratnia, N., and Serrano, J. (2014). Underwater acoustic wireless sensor networks: Advances and future trends in physical, MAC and routing layers. *Sensors* 14, 795-833. doi: 10.3390/s140100795

Ejbye-Ernst, R., Michaelsen, T. Y., Tirsgaard, B., Wilson, J. M., Jensen, L. F., Steffensen, J. F., et al. (2016). Partitioning the metabolic scope: the importance of anaerobic metabolism and implications for the oxygen- and capacity-limited thermal tolerance (OCLTT) hypothesis. *Conserv. Physiol.* 4:13. doi: 10.1093/conphys/cow019

Føre, M., Frank, K., Norton, T., Svendsen, E., Alfredsen, J. A., Dempster, T., et al. (2017). Precision fish farming: A new framework to improve production in aquaculture. *Biosystems Engineering* 173, 176-193. doi: 10.1016/j.biosystemseng.2017.10.014

Gleiss, A. C., Dale, J. J., Holland, K. N., and Wilson, R. P. (2010). Accelerating estimates of activity-specific metabolic rate in fishes: testing the applicability of acceleration data-loggers.

*Journal of Experimental Marine Biology and Ecology* 385, 85-91. doi: 10.1016/j.jembe.2010.01.012

Gruber, N. (2011). Warming up, turning sour, losing breath: ocean biogeochemistry under global change. *Phil. Trans. R. Soc. A.* 369, 1980-1996. doi: 10.1098/rsta.2011.0003

Hämäläinen, W., Järvinen, M., Martiskainen, P. and Mononen, J. (2011). Jerk-based feature extraction for robust activity recognition from acceleration data. *11th International Conference on Intelligent Systems Design and Applications (ISDA)* 831. doi: 10.1109/isda.2011.6121760

Hinch, S. G., Cooke, S, J., Healey, M. C., and Farrell, A. P. (2005). "Behavioural physiology of fish migrations: salmon as a model approach", in *Behaviour and physiology of fish. Fish Physiology Series. Volume 24*, ed. K. Sloman, S. Balshine, R. Wilson (London, UK: Academic Press), 240-285.

Horie, J., Mitamura, H., Ina, Y., Mashino, Y., Noda, T., Moriya, K., et al. (2017). Development of a method for classifying and transmitting high-resolution feeding behavior of fish using an acceleration pinger. *Animal Biotelemetry* 5, 12. doi: 10.1186/s40317-017-0127-x

Hussey, N. E., Kessel, S. T., Aarestrup, K., Cooke, S. J., Cowley, P. D., Fisk, A. T., et al. (2015). Aquatic animal telemetry: A panoramic window into the underwater world. *Science* 348:1255642. doi: 10.1126/science.1255642

Ivanov, S., Bhargava, K., and Donnelly, W. (2015). Precision farming: sensor analytics. *IEEE Intelligent Systems* 30, 76-80. doi: 10.1109/mis.2015.67

Ivy, C. M., Robertson, C. E., and Bernier, N. J. (2017). Acute embryonic anoxia exposure favours the development of a dominant and aggressive phenotype in adult zebrafish. *Proc. R. Soc. B.* 284: 20161868. doi: 10.1098/rspb.2016.1868

Jepsen, N., Schreck, C., Clements, S, and Thorstad, E. (2004) A brief discussion on the 2% tag/body mass rule of thumb. *Aquatic Telemetry: Advances and Applications – Proceedings of the Fifth Conference on Fish Telemetry*, 255-259.

Kieffer, J. D. (2000). Limits to exhaustive exercise in fish. *Comp. Biochem. Physiol. A* 126: 161-179. doi: 10.1016/S1095-6433(00)00202-6

Kolarevic, J., Aas-Hansen, Ø., Espmark, Å., Baeverfjord, G., Terjesen, B. F., & Damsgård, B. (2016). The use of acoustic acceleration transmitter tags for monitoring of Atlantic salmon swimming activity in recirculating aquaculture systems (RAS). *Aquacultural Engineering*, 72 30-39. doi: 10.1016/j.aquaeng.2016.03.002

Lau, J. H. (2016). Recent advances and new trends in Flip Chip technology. *Journal of Electronic Packaging* 138:030802-2. doi: 10.1115/1.4034037

Lee, G. G., Farrell, A. P., Lotto, A., Hinch, S. G., and Healey, M. C. (2003). Excess post-exercise oxygen consumption in adult sockeye (*Oncorhynchus nerka*) and coho (*O. kisutch*) salmon following critical speed swimming. *Journal of Experimental Biology* 206, 3253-3260. doi: 10.1242/jeb.00548

Lloret, J., Sendra, S., Ardid, M., and Rodrigues, J. J. P. C. (2012) Underwater wireless sensor communications in the 2.4 GHz ISM frequency band. *Sensors* 12, 4237-4264. doi: 10.3390/s120404237.

Makiguchi, Y., and Kojima, T. (2017). Short term effects of relative tag size and surgical implantation on feeding behaviour, survival rate, plasma lactate and growth rate in juvenile to adult rainbow trout (*Oncorhynchus mykiss*). *Fisheries Research* 185, 54-61. doi: 10.1016/j.fishres.2016.09.035

Marras, S., Killen, S. S., Domenici, P., Claireaux, G., and McKenzie, D. J. (2013). Relationships among traits of aerobic and anaerobic swimming performance in individual European sea bass *Dicentrarchus labrax*. *PLoS ONE* 8, e72815. doi: 10.1371/journal.pone.0072815

Martos-Sitcha, J. A., Bermejo-Nogales, A., Calduch-Giner, J. A., and Pérez-Sánchez, J. (2017). Gene expression profiling of whole blood cells support a more efficient mitochondrial respiration in hypoxia-challenged gilthead sea bream (*Sparus aurata*). *Frontiers in Zoology* 14: 34. doi: 10.1186/s12983-017-0220-2

Martos-Sitcha, J. A., Simó-Mirabet, P., De las Heras, V., Calduch-Giner, J. A., and Pérez-Sánchez, J. (2019). How tissue-specific contributions cope the gilthead sea bream resilience to moderate hipoxia and high stocking density. *Frontiers in Physiology* (submitted).

Martos-Sitcha, J. A., Simó-Mirabet, P., Piazzon, M. C., de las Heras, V., Calduch-Giner, J. A., Puyalto, M., et al. (2018b). Dietary sodium heptanoate helps to improve feed efficiency, growth hormone status and swimming performance in gilthead sea bream (*Sparus aurata*). *Aquaculture Nutrition* 24, 1638-1651. doi: 10.1111/anu.12799

Metcalfe, J. D., Wright, S., Tudorache, C., and Wilson, R. P. (2016). Recent advances in telemetry for estimating the energy metabolism of wild fishes. *Journal of Fish Biology* 88, 284-297. doi: 10.1111/jfb.12804

Meyer, T., Ofner, G., Bradl, S., Brunnbauer, M., and Hagen, R. (2008). Embedded wafer level ball grid array (eWLB). *Proceedings of the 2008 electronic packaging technology conference (EPTC)*, 994-998.

Neethirajan, S., Tuteja, S. K., Huang, S.-T., and Kelton, D. (2017). Recent advancement in biosensors technology for animal and livestock health management. *Biosensors and Bioelectronics* 98, 398-407. doi: 10.1016/j.bios.2017.07.015

Nelson, J. A. (1989). Critical swimming speeds of yellow perch *Perca fluvescens*: comparison of populations from a naturally acidic lake and neutral water. *Journal of Experimental Biology* 145, 239-254.

Norin, T., and Clark, T. D. (2016). Measurement and relevance of maximum metabolic rate in fish. *J. Fish. Biol.* 88, 122-151. doi: 10.1111/jfb.12796

Palmeiro, A., Martín, M., Crowther, I., and Rhodes, M. (2011). Underwater radio frequency communications. *Proceedings of the OCEANS 2011 IEEE*, 1-8. doi: 10.1109/Oceans-Spain.2011.6003580

Payne, N. L., Taylor, M. D., Watanabe, Y. Y., and Semmens, J. M. (2014). From physiology to physics: are we recognizing the flexibility of biologging tools? *Journal of Experimental Biology* 217, 317-322. doi: 10.1242/jeb.093922

Peake, S. J., (2008). Gait transition as an alternate measure of maximum aerobic capacity in fishes. *J. Fish. Biol.* 72, 645-655. doi: 10.1111/j.1095-8649.2007.01753.x

Peake, S. J., and Farrell, A. P. (2004). Locomotory behaviour and post-exercise physiology in relation to swimming speed, gait transition and metabolism in free-swimming smallmouth bass (*Micropterus dolomieu*). *J. Exp. Biol.* 207, 1563-1575. doi: 10.1242/jeb.00927

Peixoto, M. J., Salas-Leitón, E., Brito, F., Pereira, L. F., Svendsen, J. C., Baptista, T., et al. (2017). Effects of dietary *Gracilaria* sp. and *Alaria* sp. supplementation on growth performance, metabolic rates and health in meagre (*Argyrosomus regius*) subjected to pathogen infection. *Journal of Applied Phycology* 29, 433-447. doi: 10.1007/s10811-016-0917-1

Pichavant, K., Person-Le-Ruyet, J., Le Bayon, N., Severe, A., Le Roux, A. and Boeuf, G. (2001). Comparative effects of long-term hypoxia on growth, feeding and oxygen consumption in juvenile turbot and European sea bass. *Journal of Fish Biology* 59, 875-883. doi: 10.1006/jfbi.2001.1702

Plaut, I. (2001). Critical swimming speed: its ecological relevance. *Comp. Biochem. Physiol. A* 131, 41-50. doi: 10.1016/S1095-6433(01)00462-7

Remen, M., Oppedal, F., Torgersen, T., Imsland, A. K., and Olsen, R. E. (2012). Effects of cyclic environmental hypoxia on physiology and feed intake of post-smolt Atlantic salmon: Initial responses and acclimation. *Aquaculture* 326-329, 148-155. doi: 10.1016/j.aquaculture.2011.11.036

Remen, M., Oppedal, F., Imsland, A. K., Olsen, R. E., and Torgersen, T. (2013). Hypoxia tolerance thresholds for post-smolt Atlantic salmon: dependency of temperature and hypoxia acclimation. Aquaculture 416-417, 41−47. doi: 10.1016/j.aquaculture.2013.08.024

Remen, M., Nederlof, M. A. J., Folkedal, O., Thorsheim, G., Sitjà-Bobadilla, A., Pérez-Sánchez, J., et al. (2015). Effect of temperature on the metabolism, behaviour and oxygen requirements of *Sparus aurata*. *Aquaculture Environment Interactions* 7, 115-123. doi: 10.3354/aei00141

Remen, M., Solstorm, F., Bui, S., Klebert, P., Vågseth, T., Solstorm, D., et al. (2016). Critical swimming speed in groups of Atlantic salmon *Salmo salar*. *Aquaculture Environment Interactions* 8, 659-664. doi: 10.3354/aei00207

Saberioon, M., Gholizadeh, A., Cisar, P., Pautsina, A., and Urban, J. (2017). Application of machine vision systems in aquaculture with emphasis on fish: state-of-the-art and key issues. *Reviews in Aquaculture* 9, 369-387. doi: 10.1111/raq.12143

Sänger, A. M., and Stoiber, W. (2001). "Muscle fiber diversity and plasticity" in *Fish Physiology: Muscle development and growth*, ed. I. A. Johnston (San Diego, CA: Academic Press), 187-250.

Simbeye, D. S., Zhao, J., and Yang, S. (2014). Design and deployment of wireless sensor networks for aquaculture monitoring and control based on virtual instruments. *Computers and Electronics in Agriculture* 102, 31-42. doi: 10.1016/j.compag.2014.01.004

Smircich, M. G. and Kelly, J. T. (2014). Extending the 2% rule: the effects of heavy internal tags on stress physiology, swimming performance, and growth in brook trout. *Animal Biotelemetry* 2: 16. doi: 10.1186/2050-3385-2-16

Spitz, J., Chouvelon, T., Cardinaud, M., Kostecki, C., and Lorance P. (2013). Prey preferences of adult sea bass *Dicentrarchus labrax* in the northeastern Atlantic: implications for bycatch of common dolphin *Delphinus delphis*. *ICES J. Mar. Sci.* 70, 452–461. doi: 10.1093/icesjms/fss200

Staaks, G., Baganz, D., Jauernig, O., Brockmann, C., and Balzer, U. (2010) "The FischFITMonitor - A New System for Monitoring Multiple Physiological and Behavioural Parameters in Fish", in *Proceedings of Measuring Behavior 2010*, eds. A. J. Spink, F. Grieco, O. E. Krips, L. W. S. Loijens, L. P. J. J. Noldus, and P. H. Zimmerman, 433-435.

Svendsen, J. C., Tirsgaard, B., Cordero, G. A., and Steffensen, J. F. (2015). Intraspecific variation in aerobic and anaerobic locomotion: gilthead sea bream (*Sparus aurata*) and Trinidadian guppy (*Poecilia reticulata*) do not exhibit a trade-off between maximum sustained swimming speed and minimum cost of transport. *Frontiers in Physiology* 6, 6. doi: 10.3389/fphys.2015.00043

Tamayo, J., Kosaka, P. M., Ruz, J. J., San Paulo, Á., and Calleja, M. (2013). Biosensors based on nanomechanical systems. *Chemical Society Reviews* 42, 1287-1311. doi: 10.1039/C2CS35293A

Tamsin, M. (2015). Wearable biosensor technologies. *International Journal of Innovation and Scientific Research* 13, 697-703.

Teulier, L., Omlin, T., and Weber, J.-M. (2013). Lactate kinetics of rainbow trout during graded exercise: Do catheters affect the cost of transport? *Journal of Experimental Biology* 216, 4549-4556. doi: 10.1242/jeb.091058

Wilson, A. D. M., Wikelski, M., Wilson, R. P., and Cooke, S. J. (2015). Utility of biological sensor tags in animal conservation. *Conservation Biology* 29, 1065-1075. doi: 10.1111/cobi.12486

Winter, J. D. (1996). "Advances in underwater biotelemetry", in *Fisheries Techniques*, eds B. R. Murphy and D. W. Willis (Bethesda, MD: American Fisheries Society), 555-590.

Wright, S., Metcalfe, J. D., Hetherington, S., and Wilson, R. (2014). Estimating activity-specific energy expenditure in a teleost fish, using accelerometer loggers. *Marine Ecology Progress Series* 496, 19-32. doi: 10.3354/meps10528

**Figure 1.**

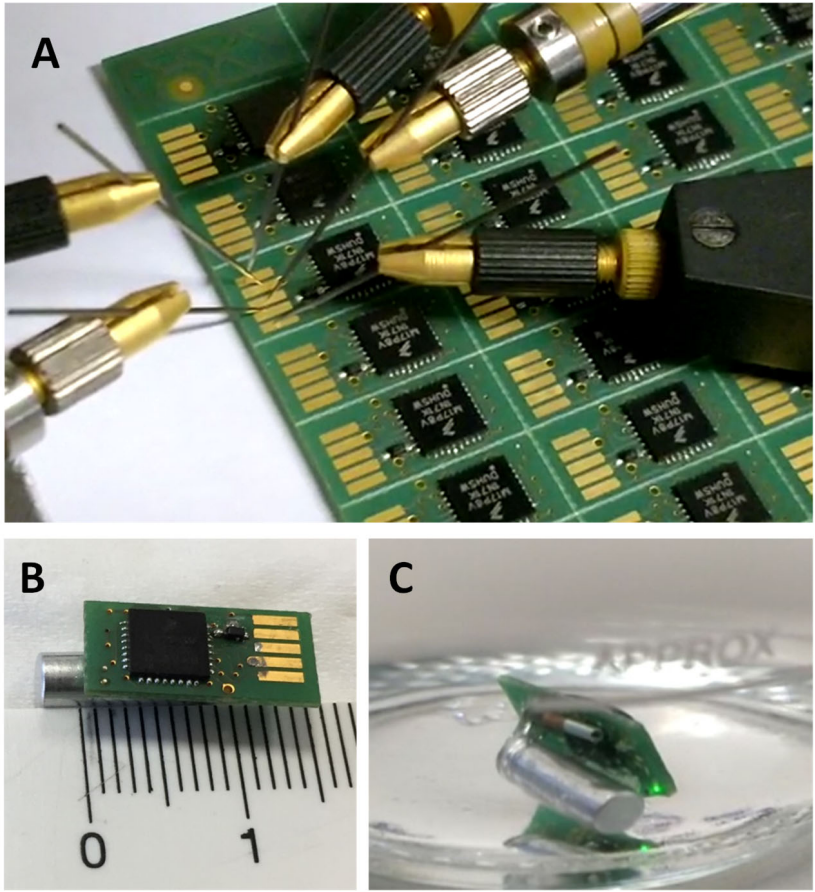

**Figure 2.**

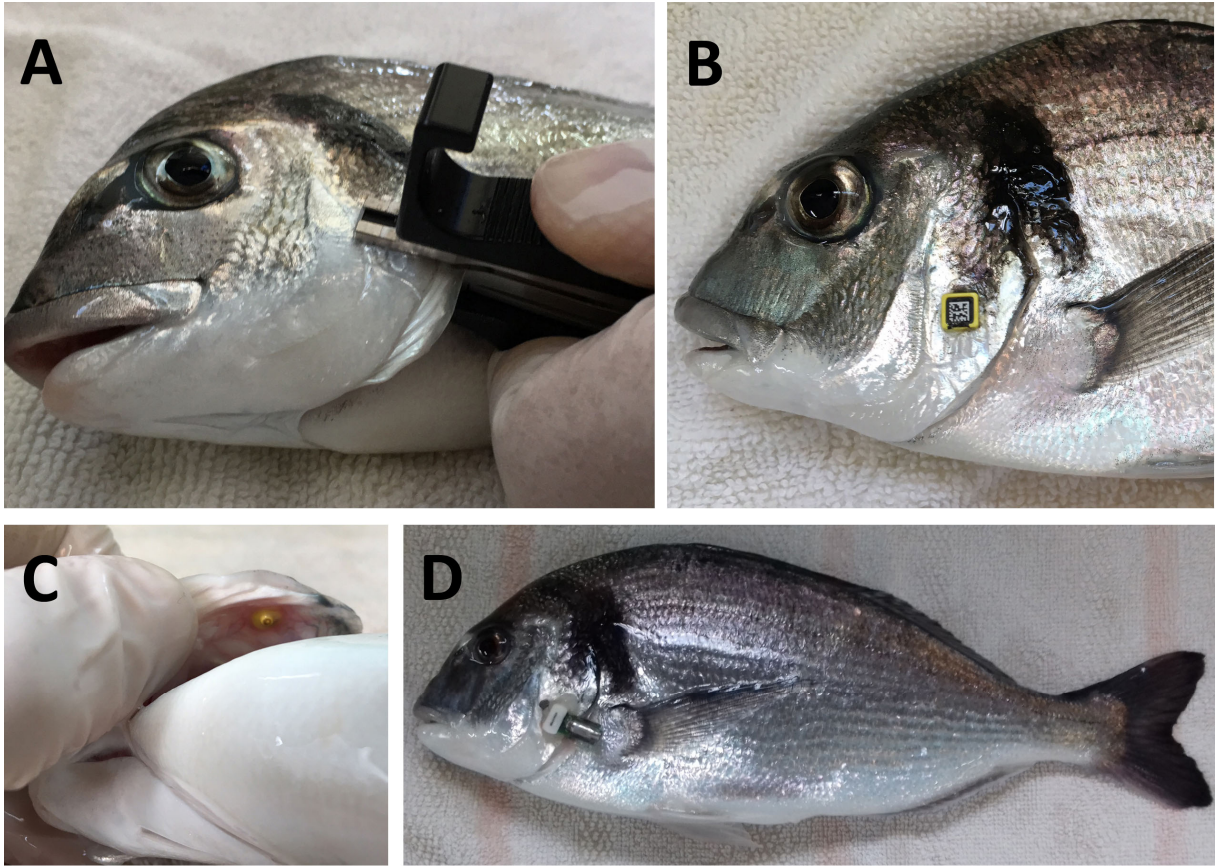

**Figure 3.**

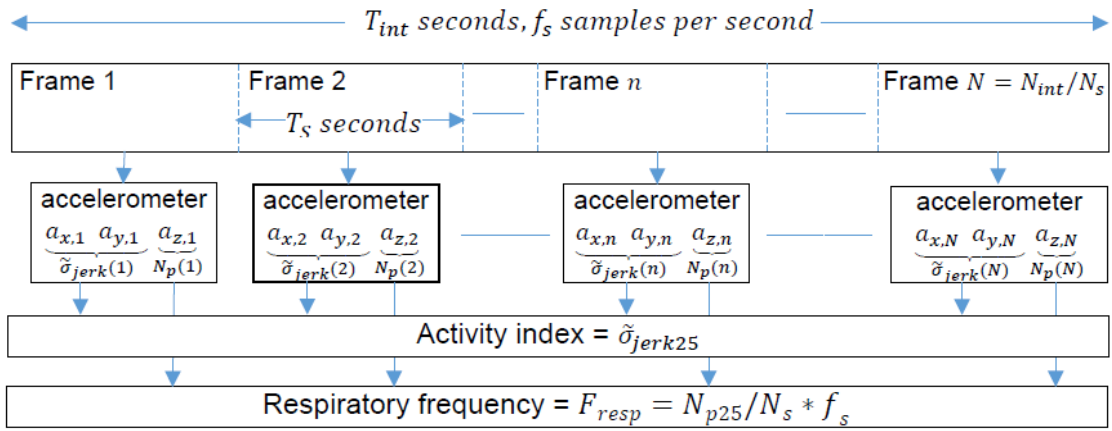

**Figure 4.**

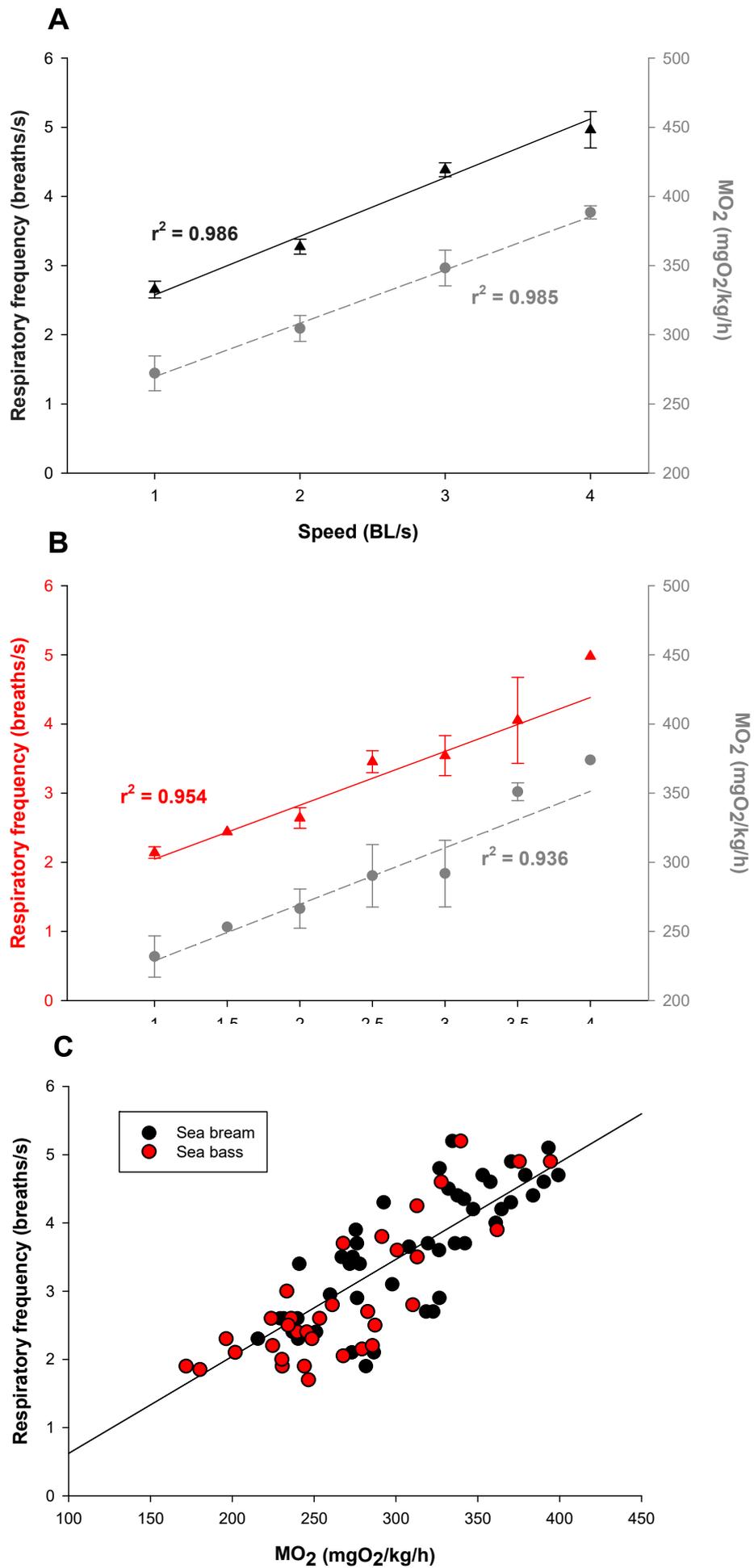

**Figure 5.**

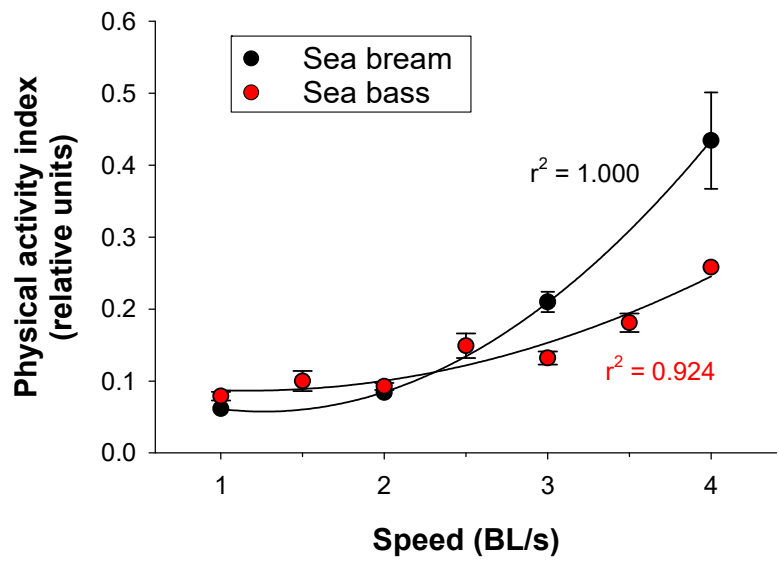

**Figure 6.**

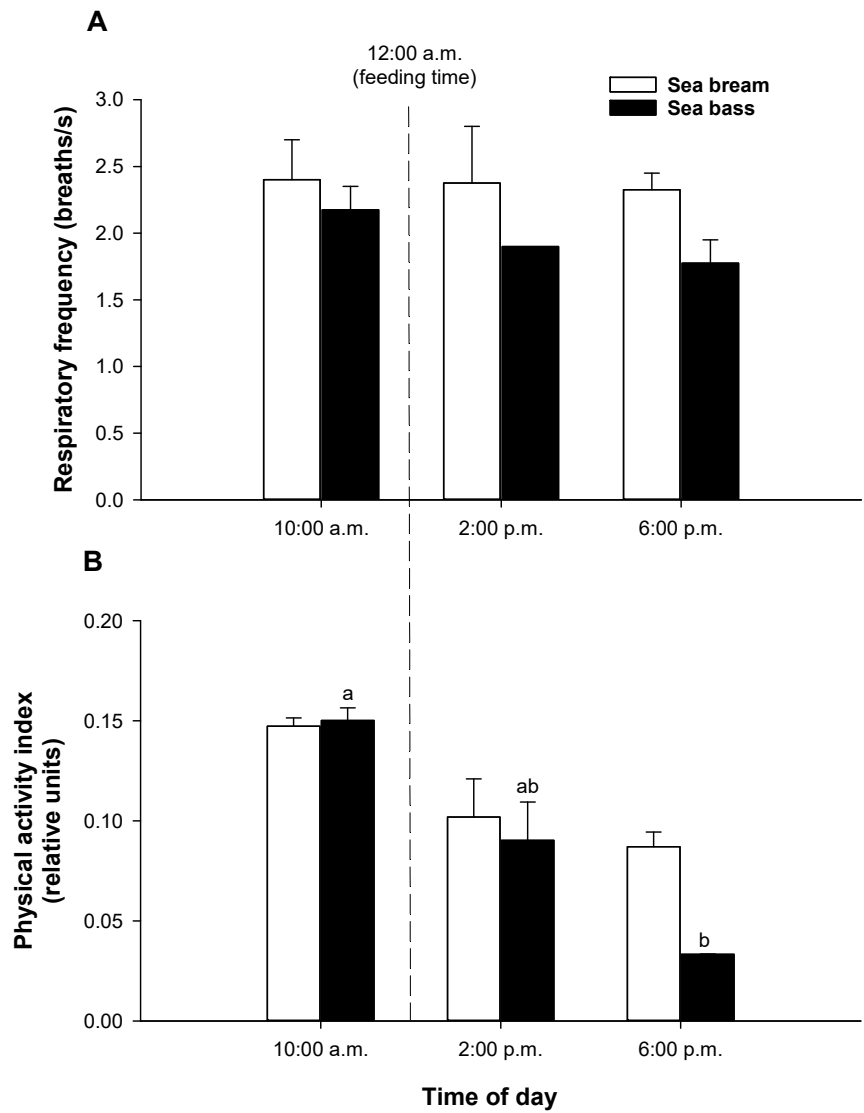

**Figure 7.**

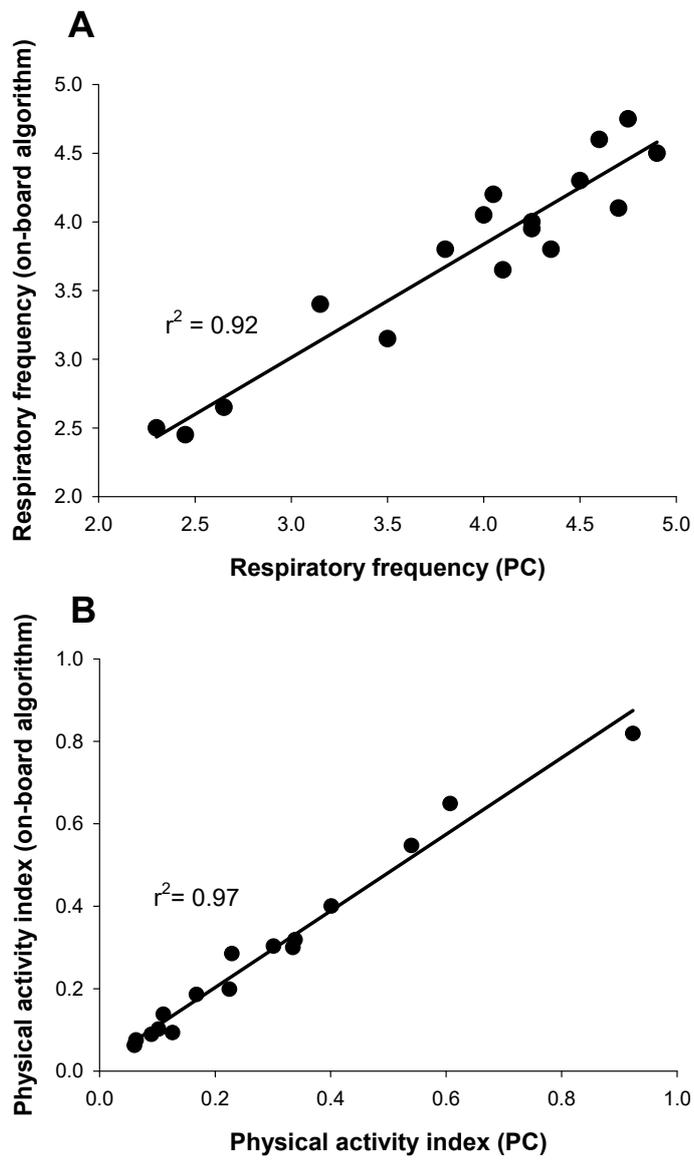

**Figure 8.**

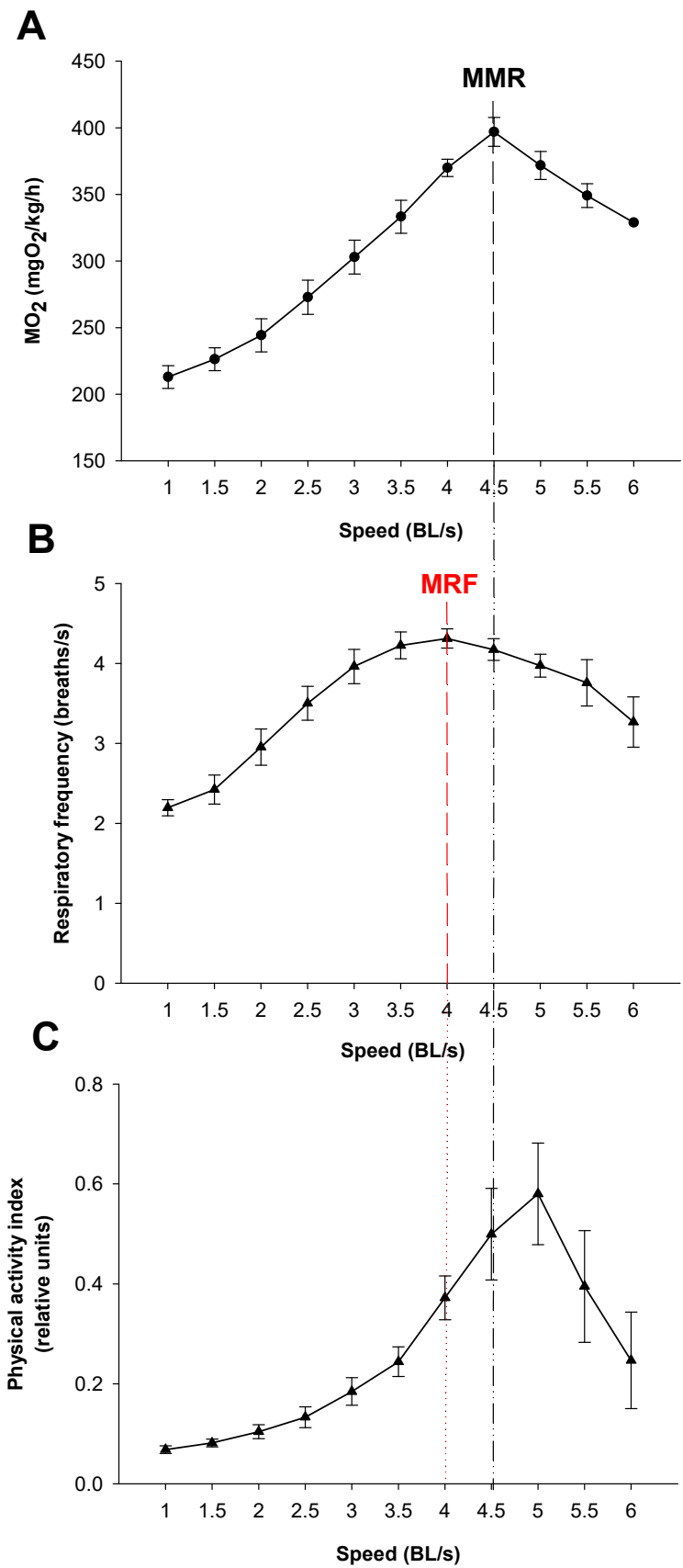

**Figure 9.**

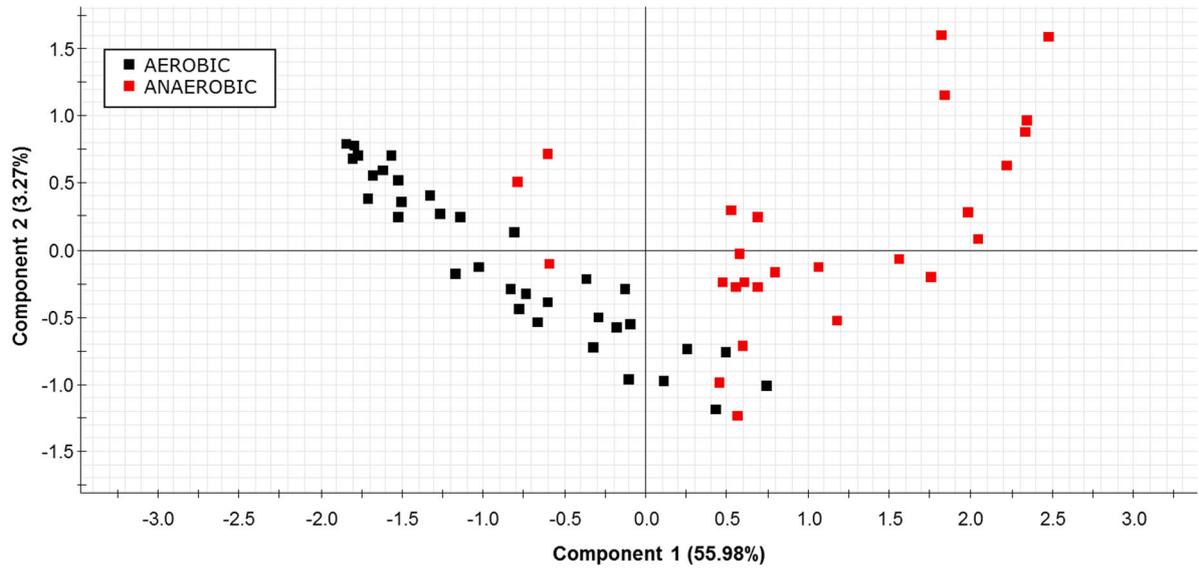

**Figure 10.**

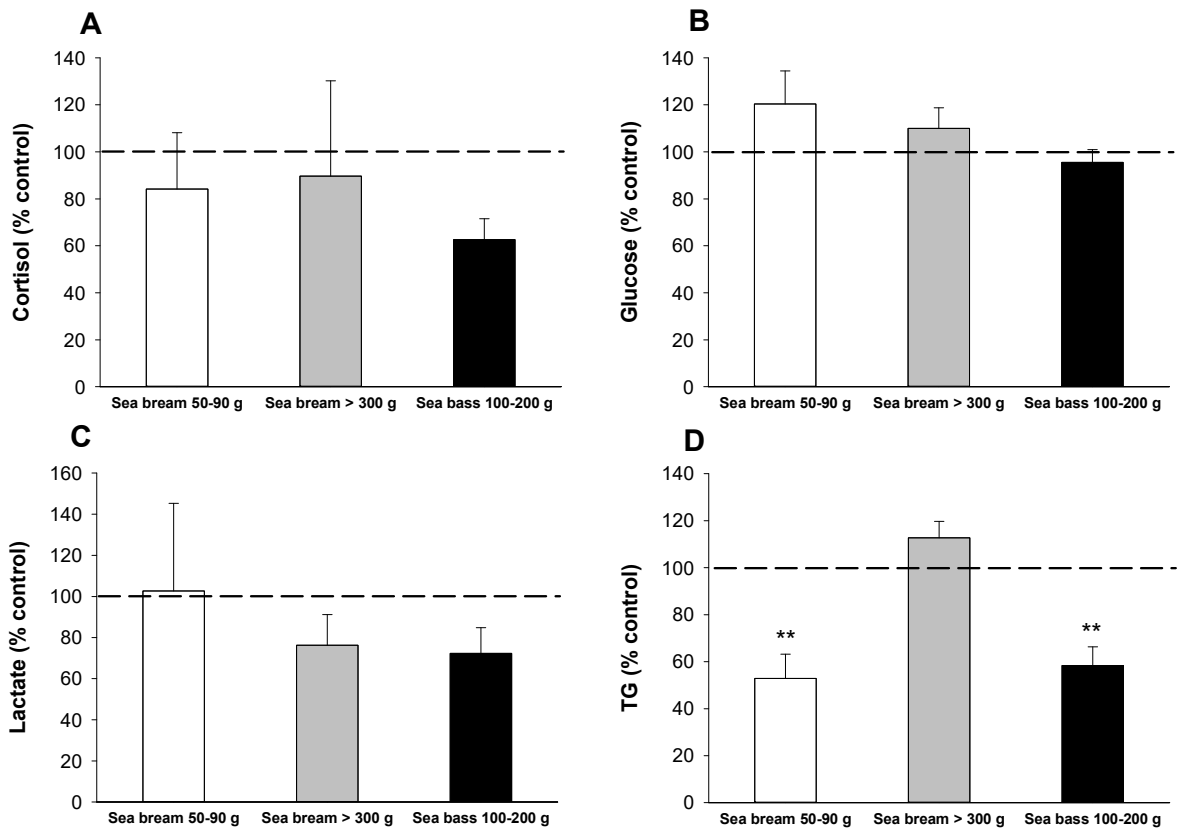

**Figure 11.**

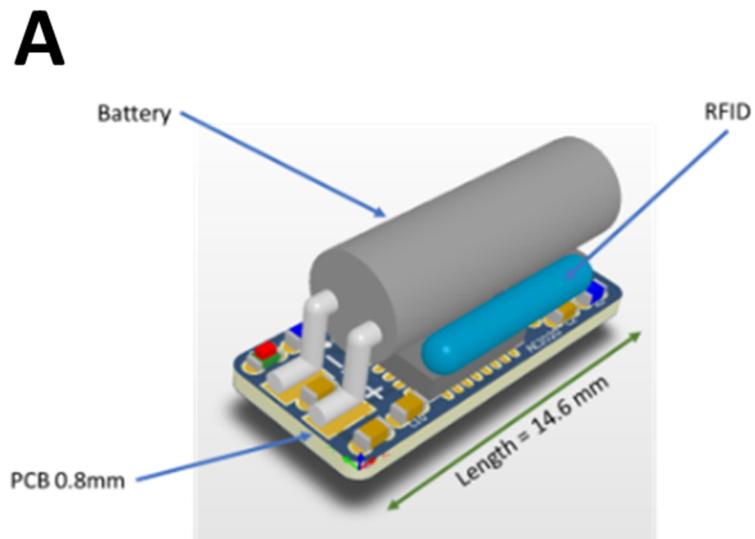

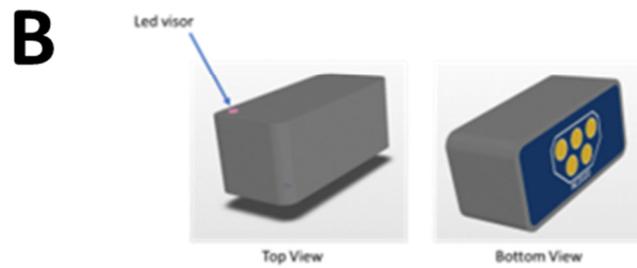